\documentclass[aps,pra,preprint,,tightenlines, groupedaddress,amssymb,amsmath,amsfont]{revtex4}

\usepackage{bm}
\usepackage{graphicx}

\usepackage{color}
\usepackage[colorlinks=true,urlcolor=blue,linkcolor=cyan,citecolor=red]{hyperref}

\begin{document}


\title{Electrodynamics of a Magnet Moving through a Conducting Pipe}


\author{M. Hossein Partovi}
\email[Electronic address: ]{hpartovi@csus.edu}
\author{Eliza J. Morris}
\email[Electronic address: ]{eliza.morris@gmail.com}
\affiliation{Department of Physics and Astronomy, California State University,
Sacramento, California 95819-6041}


\date{\today}

\begin{abstract}

The popular demonstration involving a permanent magnet falling
through a conducting pipe is treated as an axially symmetric
boundary value problem.  Specifically, Maxwell's equations are
solved for an axially symmetric magnet moving coaxially inside an
infinitely long, conducting cylindrical shell of arbitrary thickness
at nonrelativistic speeds.  Analytic solutions for the fields are
developed and used to derive the resulting drag force acting on the
magnet in integral form.  This treatment represents a significant
improvement over existing models which idealize the problem as a
point dipole moving slowly inside a pipe of negligible thickness. It
also provides a rigorous study of eddy currents under a broad range
of conditions, and can be used for precision magnetic braking
applications. The case of a uniformly magnetized cylindrical magnet
is considered in detail, and a comprehensive analytical and
numerical study of the properties of the drag force is presented for
this geometry. Various limiting cases of interest involving the
shape and speed of the magnet and the full range of conductivity and
magnetic behavior of the pipe material are investigated and
corresponding asymptotic formulas are developed.
\end{abstract}


\maketitle



\section{\label{intro} Introduction}

\subsection{\label{introA} Background and Significance}
This work is based on the popular experiment which demonstrates
magnetic damping by means of a permanent magnet falling through a
conducting pipe \cite{clack, saslow, mac, hahn}. Such an arrangement
has long been a favorite for demonstrating such topics as Faraday's
law of induction, Lenz's law, eddy currents, inductive heating and
magnetic damping.  The underlying physical process is induction
heating: the moving magnet creates a changing magnetic flux in its
vicinity which induces circulating eddy currents within the pipe
wall, thereby causing ohmic dissipation and generating a drag force
on the magnet by virtue of energy conservation. The force itself is
of course manifested as the action of the magnetic field generated
by the eddies on the permanent magnet, with its retarding nature
understood as a manifestation of Lenz's law. A particularly vivid
picture of this mechanism emerges if one views the magnet as an
assembly of circulating atomic currents moving through the pipe.
Lenz's law then implies that the induced eddies in the pipe wall
counter-circulate ahead of the moving magnet and co-circulate behind
it.  But this implies that the moving magnet is repelled in front
and attracted in rear, hence acted upon by a retarding force.

The main contribution to the literature on this subject is Saslow's
paper \cite{saslow} in which he treated the problem of eddy currents
in thin conducting sheets quite generally, and as an example
provided an approximate calculation of the drag force on a magnet
falling inside a conducting pipe. MacLatchy {\it et al.} \cite{mac}
used other methods to derive Saslow's result for the terminal speed
and pointed out a sizable discrepancy between the predictions of
this result and their measured values.  These authors introduced a
numerical modeling of the magnet which significantly improved the
agreement.  Following Saslow's calculation \cite{saslow}, analytical
treatments of the magnet-pipe problem have considered the magnet as
a point dipole moving at low speeds and the pipe as infinitely
thin-walled and long.  These assumptions imply that the only
significant length parameter in the problem is the interior diameter
of the pipe, and lead to a simple expression for the drag force. We
will refer to this limit as the {\it idealized model} and derive it
as a limiting case of our general solution in appendix \ref{idl}.
The main sources of inaccuracy in this model are the point-dipole
and thin-wall assumptions, with the low-speed approximation a
potential source of inaccuracy as well.

\subsection{\label{introB} Objectives and Limitations}

Our objective in this work is to develop a rigorous formulation of
the magnet-pipe system that avoids the approximations of the
idealized model.  To that end, we treat the case of an axially
symmetric permanent magnet moving coaxially inside an infinitely
long, conducting cylindrical shell of arbitrary thickness.  By an
\textit{axially symmetric} magnet we mean a permanent magnet whose
magnetization vector has an axially (or azimuthally) symmetric
magnitude and a uniform direction parallel to the symmetry axis.
Since any practical realization of this model will likely involve
magnet speeds far smaller than the speed of light, we will restrict
our treatment to nonrelativistic speeds, $v/c \ll 1$, where $v$ is
the magnet speed and $c$ is the speed of light. On the other hand,
we are including the possibility of the magnet being projected into
the pipe thus allowing much higher speeds than can be attained by a
magnet that starts to fall from rest under gravity. The restriction
to nonrelativistic speeds implies that we are dealing with
quasi-static sources and fields where displacement currents can be
neglected \cite{jackson}. This is so not only in the interior of the
pipe where the inclusion of the displacement currents would amount
to a minute correction of the order of ${(v/c)}^{2}$, but also
within the pipe wall where conduction currents dominate displacement
currents. To provide a basis for the latter assertion, we note that
the basic time scale generated by the motion of the magnet is of the
order of $R_1/v$, where $R_1$ is the inner radius of the pipe. This
time scale corresponds to a frequency of the order of $v/R_1$, which
implies that the ratio of displacement to conduction currents is of
the order of ${\epsilon}_{0} v/ \sigma R_1 $ where $\sigma$ is the
conductivity of the pipe wall \cite{si}. For common metals and with
$1 \leq R_1 \leq 10$ cm, this ratio is in the
$3({10}^{-7}-{10}^{-10})v/c$ range.  Since $v/c \ll 1$ by
assumption, the ratio in question is seen to be vanishingly small.

The assumption of an infinitely long pipe is unavoidable if a
manageable solution is desired. The error resulting from this
assumption, on the other hand, is small if the magnet is not close
to the pipe ends. To provide a basis for this assertion, one can use
the idealized solution, appendix \ref{idl}, to estimate the
dissipated power within the pipe segment that extends from its
actual end to its idealized end (which is infinitely far). The ratio
of this quantity to the total dissipated power is then an estimate
of the leading finite-length correction to the drag force. This
ratio is found to be of the order of ${({R}_{1}/D)}^{7}$, where $D$
is the distance from the magnet to the near end of the pipe and
$R_1$ is the pipe's interior diameter. For $R_1 /D \lesssim 1/4$,
for example, the expected correction would be of order ${10}^{-4}$
which is quite small as claimed.  Similarly, the quasi-static
approximation implies that the radiated power from the magnet-pipe
assembly is negligible.  To get an idea of the magnitude of such
radiation, consider a permanent magnet with a magnetic dipole moment
$\mathbf{m}$ moving longitudinally (i.e., parallel to $\mathbf{m}$)
through free space at nonrelativistic speeds. The radiated power
from such a source can be calculated and is found to be
${\mu}_{0}{m}^{2} {\ddot{v}}^2 / 30 \pi {c}^{5}$ where ${\dot{v}}$
and ${\ddot{v}}$ are the first and second time derivatives of the
magnet's speed. This quantity should be compared to the ohmic
dissipation in the pipe wall which for this purpose may be estimated
using the idealized model. Using the solution to the idealized model
given in appendix \ref{idl}, we estimate the ohmic dissipation rate
to be $45 {\mu}_{0}^{2} {m}^{2}{v}^{2}{\sigma} s/1024 {R_1}^{4}$
where $s$ is the thickness of the pipe wall.  For reasonable values
of the parameters, the radiated power is totally dominated by the
ohmic dissipation rate hence confirming the expectation that
radiation is completely negligible in this problem.

The case of a uniformly magnetized cylindrical magnet is considered
in detail, and a comprehensive analytical and numerical study of the
properties of the drag force is presented for this case. Various
limiting cases of interest are explored and appropriate asymptotic
formulas are developed.  The results reported here are supplemented
with a computer program posted on the web \cite{mathematica} which
can be used to compute the drag force for this case.

The rest of this paper is organized as follows: section
\ref{magnetA} deals with the characterization of the source currents
of a permanent cylindrical magnet in axial motion, and \ref{magnetB}
with the vector potential of such a magnet in the quasi-static
limit.  The electromagnetic fields of the magnet-pipe system are
found in section \ref{mag-pipe} and the magnetic drag force is
derived \ref{force}.  Section \ref{dragfA} presents the results of
numerical computations of the drag for the case of a uniformly
magnetized cylinder as well as a detailed discussion of its
dependence on magnet shape and speed, and also on the material
properties of the pipe. Concluding remarks are presented in section
\ref{concl}. Several limiting cases of the drag force are considered
in the appendices and corresponding asymptotic formulas are derived.

\section{\label{magnet} Sources and Fields of a magnet moving in free
space} The first step in our analysis is the calculation of the
electromagnetic fields of an axially symmetric permanent magnet
moving along its axis of symmetry in free space as input for the
magnet-pipe configuration.

\subsection{\label{magnetA} Source currents of a moving magnet}

By a permanent (or \textit{hard}) magnet is meant a ferromagnetic
material whose magnetization does not change when immersed in
external fields, electromagnetic or gravitational. In practice, this
requirement is met for moderate electromagnetic or gravitational
fields. Since, according to the equivalence principle, the physical
effects of acceleration are locally indistinguishable from those of
gravity \cite{weinberg}, we see that a permanent magnet is
unaffected by (moderate) acceleration.  This conclusion allows us to
characterize the accelerating magnet by means of equivalent sources
in a reference frame $\mathcal{S'}$ (with cylindrical space
coordinates $\rho', \phi', z'$) in which it is instantaneously at
rest, then transform this to the laboratory frame $\mathcal{S}$
(with coordinates $\rho, \phi, z$). In the laboratory frame, the
origin of $\mathcal{S'}$ is specified by the coordinates
$[0,0,{z}_{M}(t)]$, where ${z}_{M}(t)$ is the z-coordinate of the
center of mass of the magnet and $\dot{z}_{M}(t)\hat{\mathbf{z}}$
its velocity. Here and throughout, a caret denotes a unit vector.

The magnetization vector of an axially symmetric magnet may be
represented as $\mathcal{M'} (\rho',z')=m P(\rho',z')
\hat{\mathbf{z}}'$ in its instantaneous rest frame $\mathcal{S'}$,
where $\mathcal{M'}$ is the magnetization, $m\hat{\mathbf{z}}'$ is
the magnetic dipole moment, and $P(\rho',z')$ is an indefinite
density function whose integral is normalized to unity. It is
important to realize that in general the magnet is in accelerated
motion, so that this equation embodies the stipulation that the
magnetization of a permanent magnet is unaffected by acceleration.
The effective (or ``bound'') current density corresponding to the
above magnetization is given by ${\mathbf{J}}_{M} {=}
\mathbf{\nabla} \times \mathcal{M}$ \cite{jackson1}. Therefore, we
find ${\mathbf{J'}}_{M}(\rho',\phi',z')=-m [{\partial
P(\rho',z')}/{\partial \rho '}] \hat{\bm{\phi'}}$ in the rest frame
of the magnet.  This quantity must now be transformed to the
laboratory frame.

At this juncture we recall that the charge and current density
together transform as a 4-vector under a Lorenz transformation
\cite{jackson2}. Here the charge density of the magnet vanishes in
its rest frame, and since the current density vector is transverse
with respect to the direction of relative motion, the charge density
in the laboratory frame must vanish as well. As a result the
transformation equations reduce to the simple result that
${\mathbf{J}}_{M}(\rho,z,\phi,t) =
{\mathbf{J'}}_{M}(\rho',\phi',z')$. Note the emergence of the time
dependence in ${\mathbf{J}}_{M}(\rho,\phi,z,t)$, which originates in
the fact that $z'$ depends on time as well as on $z$.  Indeed for
nonrelativistic speeds the coordinates transform as in $\rho=\rho',
\,\, \phi=\phi'$, and $ z'=z- {z}_{M}(t)$, and we find
\begin{equation}
{\mathbf{J}}_{M}(\rho,\phi,z)=-m \frac {\partial
p[\rho,z-{z}_{M}(t)]}{\partial \rho } \hat{\bm{\phi}}, \label{3}
\end{equation}
which expresses the effective current density of the magnet in the
laboratory frame $\mathcal{S}$.  We note here that the restriction
to nonrelativistic speeds in Eq.~(\ref{3}) can be removed by simply
setting $ z'=\gamma[z- {z}_{M}(t)]$, where $\gamma
{=}{[1-{v}^{2}/{c}^{2}]}^{-1/2}$ and $v=\dot{z}_{M}(t)$.

The case of a uniformly magnetized cylindrical magnet which will be
studied in detail later corresponds to $P(\rho,z)= \Theta (\frac{L
}{ 2}- |z |) \Theta(a-\rho)/(\pi {a}^2 L)$, where $a$ and $L$ are
the radius and length of the cylinder respectively and $\Theta$ is
the standard step function. Using Eq.~(\ref{3}), we find
\begin{equation}
{\mathbf{J}}^{uni}_{M}(\rho,\phi,z,t)=\frac{m}{ \pi {a}^2 L} \Theta
(\frac{L }{ 2}- |z-{z}_{M}(t) |) \delta(a-\rho) \hat{\bm{\phi}}
\label{4}
\end{equation}
for the effective current density of a uniformly magnetized
cylindrical magnet in the laboratory frame.

\subsection{\label{magnetB} Vector Potential of the Moving Magnet}

Our task here is the calculation of the vector potential
corresponding to the current density distribution of the moving
magnet given in Eq.~(\ref{3}).  Since this current density is
transverse (or divergenceless) and the charge density vanishes, the
Lorenz and Coulomb gauges are equivalent here, with the common gauge
condition given by $\mathbf{\nabla} \cdot {\mathbf{A}}_{M} =0$,
where ${\mathbf{A}}_{M}$ is the vector potential of the moving
magnet. The electromagnetic fields of the magnet are obtained from
$\mathbf{B} = \mathbf{\nabla} \times \mathbf{A}$ and $\mathbf{E} =
-\partial \mathbf{A} / \partial t$. The azimuthal symmetry of the
current density, on the other hand, allows us to write
${\mathbf{A}}_{M}(\rho,\phi,z,t)= {A}_{M}(\rho,z,t)
\hat{\bm{\phi}}$.  Thus, like the current density, the vector
potential has an azimuthal ($\hat{\bm{\phi}}$) component only that
does not depend on the azimuthal coordinate $\phi$. This implies
that (a) the electric field is also purely azimuthal, and (b) the
magnetic field has a radial ($\hat{\bm{\rho}}$) as well as a
longitudinal ($\hat{\mathbf{z}}$) component.  It is the radial
component of the magnetic field arising from sources external to the
magnet that exerts the drag force on the moving magnet.

Using the standard solution for the vector potential in the
quasi-static limit \cite{recall}, we find
\begin{eqnarray}
{A}_{M}(\rho,z,t)=&&\frac{{\mu}_{0}}{ 4 \pi} \int dz' d \phi' \rho'
d \rho'  {[{\rho}^{2}+ {\rho'}^{2}-2 {\rho}{\rho'} \cos(\phi -
\phi')+{(z-z')}^{2}]}^{-\frac{1 }{ 2}} \nonumber \\
&& \times {J}_{M}(\rho',z',{t}) \cos (\phi - \phi'), \label{5}
\end{eqnarray}
where we have used $\hat{\bm{\phi}} \cdot \hat{\bm{\phi'}}=\cos
(\phi - \phi')$.

Next we substitute ${J}_{M}$ from Eq.~(\ref{3}) into Eq.~(\ref{5})
while making use of the representation \cite{jackson3}
\begin{eqnarray}
&& {[{\rho}^{2}+ {\rho'}^{2}-2 {\rho}{\rho'} \cos(\phi
-\phi')+{(z-z')}^{2}]}^{-\frac{1}{ 2}} = \frac{2 }{ \pi}
{\int}_{-\infty}^{\,+\infty}
dk \exp[ik(z-z')] \nonumber \\
&&\times \{ \frac{1}{ 2} {I}_{0}(|k| \rho'){K}_{0}(|k| \rho) +
{\sum}_{n=1}^{\,\infty} \cos[n(\phi - \phi')] {I}_{n}(|k|
\rho'){K}_{n}(|k| \rho) \}, \label{6}
\end{eqnarray}
which is valid for $\rho \geq \rho'$.   For $0 \leq \rho \leq
\rho'$, $\rho$ and $\rho'$ must be interchanged in Eq.~(\ref{6}).

The above substitution leads to
\begin{equation}
{A}_{M}(\rho,z,t)=-\frac{m {\mu}_{0}}{ 2  {\pi}} \int dz'  d \rho'
 dk  \exp \{ ik[z-{z}_{M}(t)-z'] \} \frac
{\partial P(\rho',z')}{\partial \rho '} \rho' {I}_{1}(|k| \rho')
{K}_{1}(|k| \rho), \label{7}
\end{equation}
where the $k$-integral ranges over $(-\infty, +\infty)$ as before. A
more convenient form of this equation obtains if we use the Fourier
representations
\begin{equation}
P(\rho,z)= {(2\pi)}^{-1/2}
 {\int}_{-\infty}^{+\infty} dk
\exp \{ ikz \}\tilde{p}(\rho,k) \label{8}
\end{equation}
and
\begin{equation}
{A}_{M}(\rho,z,t)= {(2\pi)}^{-1/2} {\int}_{-\infty}^{+\infty} dk
\exp \{ ik[z-{z}_{M}(t)] \} {\tilde{A}}_{M}(\rho,k). \label{9}
\end{equation}
Using these representations, we find from Eq.~(\ref{7})
\begin{equation}
{\tilde{A}}_{M}(\rho,k)=-{\mu}_{0} m {\int}_{0}^{+\infty} d \rho'
\frac {\partial \tilde{p}(\rho',k)}{\partial \rho '} \rho'
{I}_{1}(|k| \rho') {K}_{1}(|k| \rho). \label{10}
\end{equation}
We recall here that for $0 \leq \rho \leq \rho'$, $\rho$ and $\rho'$
must be interchanged in Eq.~(\ref{10}).

It will prove convenient to define a function ${b}_{0}(k)$ by
setting ${\tilde{A}}_{M}(\rho,k) ={b}_{0}(k) {K}_{1}(|k| \rho)$.
Thus we have from Eq.~(\ref{10}),
\begin{equation}
{b}_{0}(k)=-{\mu}_{0} m {\int}_{0}^{+\infty} d \rho' \frac {\partial
\tilde{p}(\rho',k)}{\partial \rho '} \rho' {I}_{1}(|k| \rho').
\label{11}
\end{equation}
For the case of a uniformly magnetized cylinder, we find from
Eqs.~(\ref{3}), (\ref{4}), (\ref{8}), and (\ref{11}),
\begin{equation}
{b}_{0}^{uni}(k) {=}\frac {m {\mu}_{0}}{ 2 a {\pi}^{2}}{(2 \pi
)}^{1/2}\frac{\sin(kL/2)}{ (kL/2)} {I}_{1}(|k| a). \label{12}
\end{equation}

\section{\label{mag-pipe} Electromagnetic fields of the
magnet-pipe system}

Having developed the fields of the moving magnet in free space, we
now turn to finding the fields of the magnet-pipe system. We do this
by developing the general solution followed by the imposition of
continuity conditions.

\subsection{\label{mag-pipeA} General Solution}

Our first task here is finding the governing equations for the
vector potential in the three regions (i) $a \leq \rho \leq R_1$,
(ii) $R_1 \leq \rho \leq R_2$, and (iii) $R_2 \leq \rho $, where
$R_1$ and $R_2$ are the inner and outer radii of the pipe,
respectively \cite{region0}. We will actually treat the case of a
medium with permeability $\mu$ \cite{murel} and conductivity
$\sigma$ corresponding to region (ii). For regions (i) and (iii), we
will simply replace $\mu$ and $\sigma$ with ${\mu}_{0}$ and $0$,
respectively. Recall from our discussion in section \ref{introB}
that the fields of the magnet-pipe system can be safely calculated
in the quasi-static limit where displacement currents are neglected.
Therefore the equation obeyed by the vector potential in the Lorenz
(or radiation) gauge reduces to
\begin{equation}
({\nabla}^2 - {\mu} \sigma \partial/\partial t) \mathbf{A} =0.
\label{13}
\end{equation}
As we saw in section \ref{magnetB}, the azimuthal symmetry of the
system allows us to represent the vector potential in the form
${\mathbf{A}}(\rho,\phi,z,t)={A}(\rho,z,t) \hat{\bm{\phi}}$.
Moreover, $A(\rho,z,t)$ is conveniently represented as a Fourier
integral following the example of Eq.~(\ref{9}):
\begin{equation}
A(\rho,z,t)= {(2\pi)}^{-1/2} {\int}_{-\infty}^{+\infty} dk \exp \{
ik[z-{z}_{M}(t)] \} \tilde{A}(\rho,k). \label{14}
\end{equation}
Using this representation in Eq.~(\ref{13}), we find
\begin{equation}
(\frac{ {\partial}^2 }{ {\partial \rho}^{2} } +\frac {1 }{
\rho}\frac{ {
\partial}}{ {\partial \rho} }-\frac{1}{{\rho}^2} -k^2+ik {\mu} \sigma v)
\tilde{A}(\rho,k) =0. \label{15}
\end{equation}

The general solution of Eq.~(\ref{15}) is a linear combination of
${I}_{1}(\sqrt{\kappa^2} \rho)$ and ${K}_{1}(\sqrt{\kappa^2} \rho)$,
where $\kappa^2=k^2-i {\mu} \sigma v k$ \cite{arfken}. Here and
below we will use $\sqrt{\xi}$ to denote that root of $\xi$ which
has a positive real part when $\xi$ does.  Of course when $\xi$ is
real and positive, this notation reduces to the standard convention
whereby $\sqrt{\xi}$ stands for the positive root of $\xi$.  Note
also that ${K}_{1}(\sqrt{\kappa^2} \rho)$ is singular at $\rho =0$
and vanishes exponentially as $\rho \rightarrow \infty$, while
${I}_{1}(\sqrt{\kappa^2} \rho)$ vanishes at $\rho =0$ but diverges
exponentially as $\rho \rightarrow \infty$. Using this information,
we construct the solution to Eq.~(\ref{15}) in the three regions as
follows:
\begin{equation}
{\tilde{A}}^{(i)}(\rho,k) ={\tilde{A}}_{M}(\rho,k)+{b}_{1}(k)
{I}_{1}(|k| \rho), \label{15.1}
\end{equation}
\begin{equation}
{\tilde{A}}^{(ii)}(\rho,k) ={b}_{2}(k) {K}_{1}(\sqrt{\kappa^2} \rho)
+ {b}_{3}(k) {I}_{1}(\sqrt{\kappa^2} \rho), \label{15.2}
\end{equation}
\begin{equation}
{\tilde{A}}^{(iii)}(\rho,k)={b}_{4}(k) {K}_{1}(|k| \rho), \label{16}
\end{equation}
where, it may be recalled, ${\tilde{A}}_{M}(\rho,k)
={b}_{0}(k){K}_{1}(|k| \rho)$ is the term corresponding to the
potential of the moving magnet in free space and ${b}_{0}(k)$ is
given in Eq.~(\ref{11}). Above, ${\tilde{A}}^{(n)}(\rho,k)$
represents the solution of Eq.~(\ref{15}) in the $n$th region. Note
that we have set $\sigma =0$ for regions (i) and (iii) as
stipulated.

Using the analogy of waves, one may interpret the terms appearing in
Eqs.~(\ref{15.1}-\ref{16}) as ``reflections''and ``transmissions''
resulting from the ``incident'' term $\tilde{A}_{M}(\rho,k)$. This
source term representing the contribution of the moving magnet is
incident upon the inner surface of the pipe. The second term in
region (i) is the reflection from the inner surface of the pipe into
the interior. The two terms in region (ii) correspond to a linear
combination of the transmitted term from region (i) and the
reflected term from the outer surface of the pipe. In region (iii)
we only have the transmitted term from region (ii), since there will
be no reflection from ``the surface at infinity.'' From a
mathematical point of view, on the other hand, one starts with a
linear combination of the solutions of Eq.~(\ref{15}) in each region
and proceeds to impose the required boundary conditions. Thus in
region (i), the singular term [${K}_{1}(|k| \rho)$, singular at
$\rho=0$] is normalized to represent the contribution of the moving
magnet, while in region (iii), the singular term [${I}_{1}(|k|
\rho)$, singular at $\rho=\infty$] is excluded to ensure that the
fields vanish far from the magnet-pipe system.  Of course the
physical sources of all six terms are the (bound) magnetization
currents in the magnet and in the pipe wall as well as the
conduction currents in the pipe wall.

\subsection{\label{mag-pipeB} Continuity Conditions}

Our next task is the formulation of continuity conditions across the
two boundary surfaces, the inner and outer surfaces of the pipe
wall. We recall from above that the electric field is purely
azimuthal, while the magnetic field has radial and longitudenal
components. We will denote these ${E}_{\phi}$ ($ =\hat{\bm{\phi}}
\cdot \mathbf{E}$), ${B}_{\rho}$ ($=\hat{\bm{\rho}} \cdot
\mathbf{B}$), and ${B}_{z}$ ($=\hat{\mathbf{z}} \cdot \mathbf{B}$),
respectively. Now the boundary conditions require the continuity of
${E}_{\phi}$ (Faraday's law), ${B}_{\rho}$ (absence of magnetic
monopoles), and ${H}_{z}$ (the Amp\`{e}re-Maxwell law and absence of
surface currents) across the two boundary surfaces, where
$\mathbf{H}=\mathbf{B}/{\mu}_{0}$ in regions (i) and (iii), and
$\mathbf{H}=\mathbf{B}/{\mu}$ in region (ii). When expressed in
terms of $\mathbf{ A}$, the first two of these conditions require
the continuity of $A$ while the third condition requires the
continuity of $ {\mu}^{-1}\partial (\rho A) /\partial \rho$.  An
inspection of Eq.~(\ref{12}) shows that these conditions must also
be obeyed by $\tilde{A}$.  This gives us the continuity equations
that must be imposed on the solutions of Eq.~(\ref{13}). Recall that
we must also apply the conditions $\kappa \rightarrow k$ and $\mu
\rightarrow {\mu}_{0}$ in regions (i) and (iii).

Upon imposing the above-stated continuity conditions on the
solutions given in Eqs.~(\ref{14}-\ref{16}) at the boundary surfaces
$\rho=R_1$ and $\rho=R_2$, we find the following set of equations:
\begin{eqnarray}
{b}_{0}(k) {K}_{1}(|k| R_1)+{b}_{1}(k) {I}_{1}(|k| R_1) &=&
{b}_{2}(k) {K}_{1} (\sqrt{\kappa^2} R_1) + {b}_{3}(k)
{I}_{1}(\sqrt{\kappa^2} R_1), \nonumber
\\ {b}_{2}(k) {K}_{1} (\sqrt{\kappa^2} R_2) + {b}_{3}(k) {I}_{1}(\sqrt{\kappa^2}
R_2) &=& {b}_{4}(k) {K}_{1}(|k| R_2), \nonumber
\\\frac{|k|}{\mu_0}[ -{b}_{0}(k) {K}_{0}(|k| R_1)+{b}_{1}(k)
{I}_{0}(|k| R_1)] &=& \frac{\sqrt{\kappa^2}}{ \mu} [- {b}_{2}(k)
{K}_{0} (\sqrt{\kappa^2}R_1) + {b}_{3}(k) {I}_{0}(\sqrt{\kappa^2}
R_1)], \nonumber
\\ \frac{\sqrt{\kappa^2}}{
\mu} [ -{b}_{2}(k) {K}_{0} (\sqrt{\kappa^2} R_2)] + {b}_{3}(k)
{I}_{0}(\sqrt{\kappa^2} R_2)] &=& \frac{|k|}{\mu_0}[-{b}_{4}(k)
{K}_{0}(|k| R_2)]. \label{18}
\end{eqnarray}
This linear set can be solved by standard methods to find the
unknown coefficients ${b}_{1}(k)$ through ${b}_{4}(k)$.  To avoid
unnecessary writing, we will only record the solution for
${b}_{1}(k)$, since this is the coefficient that will be needed for
the calculation of the drag force in section \ref{dragfA}:
\begin{eqnarray}
b_1(k)&=& \{ [ K_{0}(|k| R_1) {K}_{0}(|k| R_2)T_{11} +\beta
{K}_{0}(|k| R_1) {K}_{1}(|k| R_2) T_{10} \nonumber \\&& -\beta
{K}_{1}(|k| R_1) {K}_{0}(|k| R_2)T_{01} -{\beta}^{2} {K}_{1}(|k|
R_1) {K}_{1} (|k|R_2)T_{00}] \nonumber \\&& \div [ I_{0}(|k| R_1)
{K}_{0}(|k| R_2)T_{11} +\beta {I}_{0}(|k| R_1) {K}_{1}(|k| R_2)
T_{10} \nonumber \\&& +\beta {I}_{1}(|k| R_1) {K}_{0}(|k| R_2)T_{01}
+{\beta}^{2} {I}_{1}(|k| R_1) {K}_{1} (|k| R_2)T_{00}] \} b_0(k) ,
\label{19}
\end{eqnarray}
where
\begin{eqnarray}
T_{00}&=&{K}_{0}(\alpha |k| R_1){I}_{0} (\alpha |k| R_2)
-{I}_{0}(\alpha |k| R_1){K}_{0}(\alpha |k|R_2)  \nonumber \\
T_{01}&=&{K}_{0}(\alpha |k| R_1){I}_{1}(\alpha |k| R_2)
+{I}_{0}(\alpha |k|R_1){K}_{1} (\alpha |k| R_2)\nonumber \\
T_{10}&=&{K}_{1} (\alpha |k| R_1){I}_{0}(\alpha |k| R_2)+
{I}_{1}(\alpha |k| R_1) {K}_{0}(\alpha |k| R_2) \nonumber \\
T_{11}&=&{K}_{1}(\alpha |k| R_1){I}_{1}(\alpha |k| R_2)
-{I}_{1}(\alpha |k| R_1){K}_{1} (\alpha |k| R_2), \label{20}
\end{eqnarray}
and
\begin{eqnarray}
\alpha&=&{\sqrt{\kappa^2}}/{ |k|}=\sqrt{1-i\frac{
{\mu}_{0}{\mu}_{rel}\sigma v}{k}}, \nonumber \\
\beta&=&\frac{\alpha}{{\mu}_{rel}}=\frac{1}{{\mu}_{rel}}
\sqrt{1-i\frac{ {\mu}_{0}{\mu}_{rel}\sigma v}{k}}. \label{21}
\end{eqnarray}
Here ${\mu}_{rel}={\mu}/{\mu}_0$ represents the relative
permeability of the pipe material.  It is worth noting that $\alpha$
and $\beta$ have positive real parts by construction.

With the coefficients given in Eqs.~(\ref{19}-\ref{21}), we have in
Eqs.~(\ref{11}), (\ref{14}), and (\ref{15.1}-\ref{16}) a complete
solution to the magnet-pipe system in all regions.

\section{\label{force} Drag Force on the Moving Magnet}

We are now in position to calculate the braking force exerted on the
moving magnet.  Recall that the density of magnetic force exerted at
a point $\mathbf{r}$ of a current distribution
$\mathbf{J}(\mathbf{r})$ is given by $\mathbf{J}(\mathbf{r}) \times
\mathbf{B}(\mathbf{r})$, where $\mathbf{B}(\mathbf{r})$ is the
magnetic field strength at the point in question.  Thus the force
exerted on the moving magnet is given by
\begin{equation}
\mathbf{F}=\int d^3{\mathbf{r}}\,\, \mathbf{J}_M(\mathbf{r},t)
\times \mathbf{B}(\mathbf{r}), \label{22}
\end{equation}
where $\mathbf{J}_M(\mathbf{r},t)$ is the effective current density
of the magnet given in Eq.~(\ref{3}). Note that
$\mathbf{B}(\mathbf{r})$ can be replaced with
$\mathbf{B}^{ext}(\mathbf{r})$ in Eq.~(\ref{22}), where the latter
is the magnetic field produced in region (i) by sources external to
the magnet.  Using Eqs.~(\ref{14}) and (\ref{15.1}), we find
\begin{equation}
\mathbf{B}^{ext}(\mathbf{r})=\nabla \times {(2\pi)}^{-1/2}
{\int}_{-\infty}^{+\infty} dk \exp \{ ik[z-{z}_{M}(t)] \} b_1(k)
{I}_{1}(|k| \rho)\hat{\bm{\phi}}. \label{23}
\end{equation}
Combining Eqs.~(\ref{3}), (\ref{8}), and (\ref{23}), and performing
a few straightforward operations, we find from Eq.~(\ref{22})
\begin{equation}
\mathbf{F}=-2 \pi i m \hat{\mathbf{z}} {\int}_{-\infty}^{+\infty} k
dk {\int}_{0}^{+\infty} \rho d\rho {b}_{1}(k) {I}_{1}(|k|\rho) \frac
{\partial \tilde{p}(\rho,-k)}{\partial \rho}. \label{24}
\end{equation}
At this juncture we use Eq.~(\ref{11}) to rewrite this result in the
form
\begin{equation}
\mathbf{F}=2 \pi i {\mu}_{0}^{-1} \hat{\mathbf{z}}
{\int}_{-\infty}^{+\infty} k dk {b}_{0}(-k){b}_{1}(k). \label{25}
\end{equation}

The quantity ${b}_{1}(k)$ was defined in Eqs.~(\ref{19}-\ref{21})
\textit{et seq}, and can conveniently be expressed as
$b_1(k)=Q(k)b_0(k)$.  An inspection of Eqs.~(\ref{19}-\ref{21})
shows that the real and imaginary parts of $Q(k)$ so defined are
even and odd functions of $k$, respectively.  On the other hand,
Eqs.~(\ref{11}) and (\ref{8}) show that
${b}_{0}(k)={b}^{*}_{0}(-k)$. These properties allow us to rewrite
Eq.~(\ref{25}) as
\begin{equation}
\mathbf{F}=-4 \pi {\mu}_{0}^{-1} \hat{\mathbf{v}}
{\int}_{0}^{+\infty} k dk {|{b}_{0}(k)|}^{2} \textrm{Im}[Q(k)],
\label{25.1}
\end{equation}
where we have also replaced $\hat{\mathbf{z}}$ by $\hat{\mathbf{v}}$
to make the retarding nature of the force manifest.  This equation
gives the drag force acting on the moving magnet and is a central
result of our analysis.

Using Eq.~(\ref{12}) which gives ${b}_{0}^{uni}(k)$ corresponding to
the uniformly magnetized magnet, we find from Eq.~(\ref{25.1})
\begin{equation}
\mathbf{F}^{uni}=-\hat{\mathbf{v}} \frac {\mu_0 m^2 }{2 {\pi}^2}
{\int}_{0}^{+\infty} dk k^3 {\left[ \frac{\sin(kL/2)}{ (kL/2)}
\right]}^2 {\left[ \frac{{I}_{1}(ka) }{ (ka/2)} \right]}^2
\textrm{Im}[Q(k)]. \label{25.2}
\end{equation}
This equation gives the drag force acting on the moving magnet for
the case of a uniformly magnetized cylinder.

It is instructive to rederive the drag force formula in
Eq.~(\ref{25.1}) from energy conservation. Let us consider the
magnet-pipe system at a moment the magnet is moving through the pipe
with velocity ${\mathbf{v}}$. Under the quasi-static conditions
stipulated in the section \ref{introB}, ohmic power dissipation in
the pipe wall must be balanced by a decrease in the kinetic energy
of the moving magnet, since the power flow into the electromagnetic
field configuration, including radiation, is negligible. But the
decrease in kinetic energy corresponds to a resistive force
according to the work-energy theorem, $P_{ohm}=-\mathbf{F}\cdot
\mathbf{v}$, where $P_{ohm}$ is the rate of ohmic dissipation (or
Joule heating). Needless to say, the magnet may be experiencing
non-electromagnetic forces such as gravity or air drag which would
have their own power contributions. Now Poynting's theorem assures
us that the dissipated power equals the flux of Poynting's vector
into the pipe wall through its inner surface.  Putting these two
observations together, we arrive at
\begin{equation} \mathbf{F}\cdot \mathbf{v}=-{\int}_{\rho
=R_1} \rho d \phi dz \mathbf{S} \cdot \hat{\bm{\rho}}, \label{26}
\end{equation}
where $\mathbf{S}$ is Poynting's vector, given by $={\mu}_{0}^{-1}
\mathbf{E}\times \mathbf{B}$.  Since by symmetry $\mathbf{F}$ must
have an axial direction, we can rewrite Eq.~(\ref{26}) as
\begin{equation}
\mathbf{F}=- \hat{\mathbf{v}} \frac{1}{{\mu}_{0}v} {\int}_{\rho
=R_1} \rho d \phi dz \hat{\bm{\rho}}  \cdot \left[ \frac{\partial
\mathbf{A}}{\partial t} \times (\nabla \times \mathbf{A} ) \right],
\label{27}
\end{equation}
where we have used $\mathbf{B} = \mathbf{\nabla} \times
\mathbf{A}$ and $\mathbf{E} = -\partial \mathbf{A} /
\partial t$.

At this juncture we use Eq.~(\ref{14}) to substitute the Fourier
representation of $A^{(i)}$ into Eq.~(\ref{27}).  This allows us to
perform all implied integrations except one, with the result
\begin{equation}
\mathbf{F}=-2 \pi i R_1 {\mu}_{0}^{-1}  \hat{\mathbf{v}}
{\int}_{-\infty}^{\infty}k dk { \left[ {\tilde{A}}^{(i)}(\rho,k)
\frac{\partial {\tilde{A}}^{(i)}(\rho,-k)}{\partial \rho} \right]}_
{\rho=R_1}. \label{28}
\end{equation}
Next, we use Eq.~(\ref{15.1}) to replace ${\tilde{A}}^{(i)}(\rho,k)$
with its expression in terms of modified Bessel functions.  Of the
resulting four terms inside the square brackets, two are even in $k$
and make no contribution to the integral in Eq.~(\ref{28}). The
other two terms equal $|k| b_0(k) b_1(-k)W$, where $W=1/|k|R_1$ is
the Wronskian of $K_1(|k|R_1)$ and $I_1(|k|R_1)$ \cite{arfken}. Upon
replacing $b_1(-k)$ with $b_0(-k)Q(-k)$ in Eq.~(\ref{28}), and
recalling that the real and imaginary parts of $Q(k)$ are even and
odd functions of $k$ respectively, we recover Eq.~(\ref{25.1}). This
completes the derivation of the magnetic drag force from energy
conservation.

\section{\label{dragfA} Properties of the Drag Force}

The remainder of this paper is devoted to a detailed discussion of
the properties of the drag force for the case of a uniformly
magnetized cylinder, given in Eq.~(\ref{25.2}). It should be
remembered, however, that the magnetization distribution of a
typical magnet depends on its type and manufacturing method, and
almost certainly deviates from uniformity. The following study is
thus intended to provide a benchmark that typifies general
properties.

We shall begin our study by characterizing the main features of the
drag force here, including its dependence on the shape and speed of
the magnet as well as the material properties of the pipe wall. In
the appendices, we will derive asymptotic expressions for the drag
force of Eq.~(\ref{25.2}) in a number of physically interesting
limiting cases.

\subsection{\label{shape} Dependence on Magnet Shape}

Let us start by examining the dependence of the drag force on the
dimensions of the magnet.  We have already arranged Eq.~(\ref{25.2})
in such a way as to isolate and highlight the dependence of the drag
force on the relevant parameters. The force is opposite the velocity
\cite{Qhas} and it is scaled by the square of the magnet's dipole
moment \cite{scale}. For a fixed value of the dipole moment, the
dependence of the force on the dimensions of the magnet is entirely
contained in the two bracketed factors in Eq.~(\ref{25.2}), the
first of which depends on the magnet length $L$ and the second on
its radius $a$. Each of these factors has been normalized to unity
at the point-dipole limit, the first at $L = 0$ and the second at $a
= 0$. An inspection of the second factor shows that it increases
monotonically with increasing $a$, which is expected as such an
increase brings the source currents in the magnet closer to the eddy
currents in the pipe wall thus increasing the interaction force.  On
the other hand, the envelope of the first factor clearly decreases
with increasing magnet length, suggesting a corresponding decrease
in the drag force with increasing magnet length.  This is in fact
confirmed by our numerical results, and reflects the weakening of
the external magnetic field with increasing magnet length (with a
fixed magnetic dipole moment as stipulated), which field eventually
vanishes as $L/a \rightarrow \infty$, just as it would for an ideal
solenoid. It must be remembered, however, that $a$ can at most equal
$R_1$, and $L$ must remain small compared to the distance from the
magnet to the pipe ends. It is convenient in this regard to define
an ordered pair of dimensionless parameters, $(L/2a,a/R_1)$,
characterizing the inverse aspect ratio of the magnet and the
tightness of its fit inside the pipe, respectively. Thus for a given
dipole strength, the pair $(0,1.00)$ corresponds to a wafer-shaped
magnet that would just fit inside the pipe while $(1,0.60)$
describes a ``square cylinder'' filling 36\% of the interior cross
section of the pipe.

\begin{figure}
\includegraphics[]{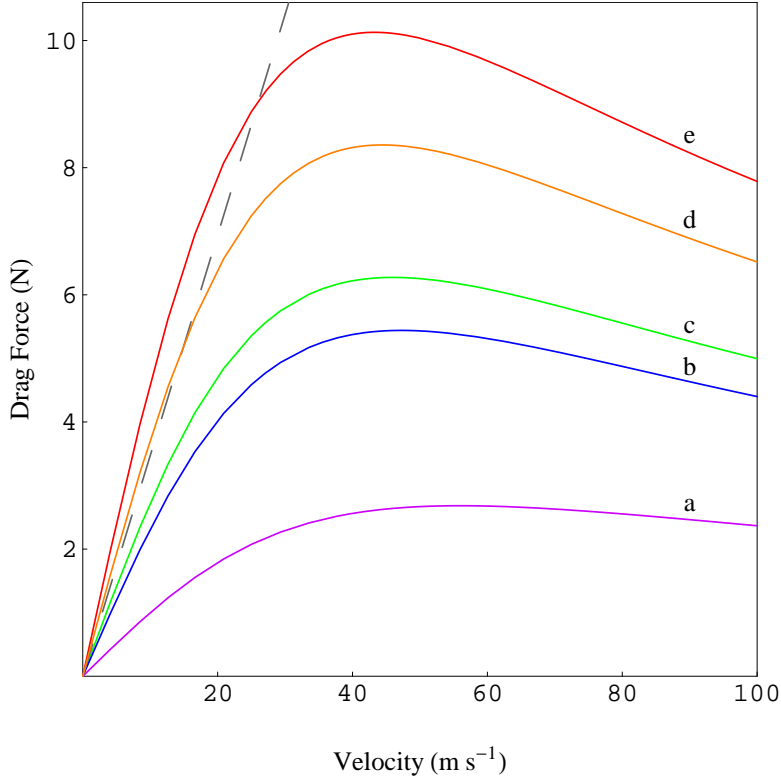}
\caption{Plot of the drag force versus the magnet speed for a fixed
value of the dipole moment and four different shape parameters
$(L/2a,a/R_1)$ (see the text for definitions): (a) typical cylinder,
($\frac{2}{1},0.60$), (b) ``square'' cylinder, ($\frac{1}{1},0.60$),
(c) ``point-like cylinder'' ($\frac{1}{1},\simeq 0$), (d) short
cylinder ($\frac{5}{8},0.96$), and (e) circular wafer, ($\simeq 0,
\frac{3}{5}$).  The dashed line represents the idealized limit.}
\label{fig1}
\end{figure}

Numerical results were obtained from Eq.~(\ref{25.2}) using a
\textit{Mathematica} program developed for this purpose
\cite{mathematica}. Figure~\ref{fig1} shows a plot of the drag force
versus magnet speed for $m = 1.00$ A ${\textrm{m}^2}$, ${\mu}_{rel}
= 1.00$, $\sigma = 5.00 \times 10^7$ ${\Omega}^{-1}$
${\textrm{m}}^{-1}$, $R_1 = 10.0$ mm, $R_2 = 11.0$ mm, and five
different shapes, (a) (typical cylinder, loose fit) $\Rightarrow$
($\frac{2}{1},0.60$), (b) (``square'' cylinder, loose fit)
$\Rightarrow$ ($\frac{1}{1},0.60$), (c) point-like cylinder
$\Rightarrow$ ($\frac{1}{1},\simeq 0$), (d) (short cylinder, snug
fit) $\Rightarrow$ ($\frac{5}{8},0.96$) (e) (circular wafer, loose
fit) $\Rightarrow$ ($\simeq 0, 0.60$). The dashed line, on the other
hand, represents the idealized model limit derived in appendix
\ref{idl}, $\mathbf{F}^{idl}=- \hat{\mathbf{v}}45 {\mu}_{0}^{2}
{m}^{2}{v}{\sigma} s/1024 {R_1}^{4}$, with the pipe thickness $s$
set equal to $R_2-R_1=1.0$ mm in order to facilitate comparison to
the exact results.

The five cases shown in Fig.~\ref{fig1} demonstrate the effect of
the shape of the magnet on the drag force.  A comparison of cases
(a) and (b) in Fig.~\ref{fig1} clearly shows that for fixed values
of the dipole moment and speed, the drag force increases as the
magnet is shortened. Comparing cases (b) and (d), or (c) and (e), on
the other hand, we see an increase in the drag force with increasing
magnet diameter. These five cases clearly demonstrate the strong
influence of the shape of the magnet on the drag force.  In
particular, they clearly demonstrate the quantitative inadequacy of
the point-dipole approximation, case (c), across a broad range of
speeds, as discussed in section \ref{introA}.

The dashed line representing the idealized model corresponds to
augmenting the point-dipole approximation by the thin-wall
assumption, $(R_2-R_1)/R_1\rightarrow 0$ \cite{thinwall}. A
comparison of the dashed line and curve (c) in Fig.~\ref{fig1} shows
a relative deviation of about $10$\% in the low-speed regime. This
deviation should be compared with the ratio $(R_2-R_1)/R_1$
characterizing the relative thickness of the pipe, which equals
$0.100$ for all cases in Fig.~\ref{fig1}. It is clear that the error
caused by the thin-wall assumption is, in general, unacceptably
large for a reasonable agreement with measurement results under
typical conditions. Exceptions can of course occur: for case (d)
corresponding to a short cylindrical magnet fitting snugly inside
the pipe, the errors caused by the point-dipole and thin-wall
approximations nearly cancel one another in the low-speed regime.

It is appropriate at this juncture to compare the prediction of our
analysis to the measured results of MacLatchy {\it et al.}
\cite{mac}, bearing in mind the important caveat that the
magnetization distribution in their experiment was not uniform.
Using the parameter values reported by these authors, namely $m =
0.67$ A ${\textrm{m}^2}$, $\sigma = 5.08 \times 10^7$
${\Omega}^{-1}$ ${\textrm{m}}^{-1}$, $R_1 = 7.29$ mm, $R_2 = 7.96$
mm, $L = 6.4$ mm, and $a = 6.3$ mm, and equating their reported
magnet (plus tape) weight of $0.060$ N to the magnetic drag force of
Eq.~(\ref{25.2}) as well as setting ${\mu}_{rel} = 1.00$ appropriate
for copper, we find a magnet speed of $11.9 \pm 0.5$ cm
${\textrm{s}}^{-1}$. MacLatchy {\it et al.} reported a measured
terminal speed of $12.7 \pm 0.4$ cm ${\textrm{s}}^{-1}$, and
compared this to the prediction of the idealized model, $17.8$ cm
${\textrm{s}}^{-1}$.  Note that in this comparison we are neglecting
air drag on the magnet, estimated by the authors to be less than
$0.1$\% of the magnetic drag force hence deemed negligible. The
quoted error of $0.5$ cm ${\textrm{s}}^{-1}$ in our calculated
result is estimated on the basis of the (implied) precision level of
the measured values given by MacLatchy {\it et al.} for $L$, $a$,
$m$, $R_2-R_1$, and the magnet weight.  Keeping in mind that the
magnetization distribution of the ``button'' magnet used in the
experiment was not uniform, and the fact that the drag force is
rather sensitive to this distribution as demonstrated by the
numerical modeling of MacLatchy {\it et al.}, we conclude that our
predicted value of the magnet speed agrees with the measured value
within the uncertainties.

The circular wafer ($L=0$) geometry merits special attention since,
in that case, the drag force grows without limit as $a \rightarrow
R_1$. In other words, the shape characterized by ($0,1$) is a
singular limit where the magnetic drag force diverges.  This
behavior may be understood as follows.  In the limit of ($0,1$)
geometry, the effective (or ``bound'') source currents of the magnet
are concentrated in a single current ring of radius $R_1$. However,
such a source would induce similarly concentrated eddy current rings
in the pipe wall on its interior surface, i.e., at \textit{zero
distance} from the source itself, thus causing infinite interaction
forces due to overlapping current rings. This situation is analogous
to the divergence of the image force on a point charge as it
approaches the surface of a conductor. It is worth remarking that
while the limit itself is unphysical, the sharp increase in the
magnetic drag force for $L \ll a$ and $a \rightarrow R_1$ is very
real and already discernible from a comparison of cases (c) and (e)
in Fig.~\ref{fig1}. In practice, this sharp increase in the magnetic
drag would be accompanied by a parallel increase in the air drag
force caused by the unavoidable constriction in the flow of air
around the magnet \cite{vac}.

The effect just discussed can also be understood in reference to
Eq.~(\ref{25.2}).  Using the asymptotic properties of the modified
Bessel functions \cite{arfken}, we find that $\textrm{Im}[Q(k)]
\rightarrow \exp (-2kR_1)/k $ as $k  \rightarrow \infty$.  Now for
$L/a \rightarrow 0$ and $a \rightarrow R_1$, the shape factors in
Eq.~(\ref{25.2}) behave like $\exp (2kR_1)/k^3$ as $k  \rightarrow
\infty$.  Putting these two statements together, we conclude that
the integrand in Eq.~(\ref{25.2}) behaves like $k^{-1}$ for large
$k$, which implies a logarithmic divergence at the upper limit of
the integral.  Recalling that $k$ is the Fourier conjugate of $z$
[cf. Eq.~(\ref{25.2})], so that large values of $k$ are associated
with short values of $z$, we conclude that this divergence is a
short-distance effect.  This is of course the conclusion we reached
above on physical grounds.

\subsection{\label{speed} Dependence on Magnet Speed}

The dependence of the drag force on the speed of the magnet is
contained in $Q(k)$ through the combination $\mu \sigma v$ and is
more involved than the dependence on its shape. It proves expedient
to discuss this dependence in terms of a length parameter defined by
$l_0={({\mu}_0 \sigma v)}^{-1}$. As discussed below, $l_0$
represents the penetration depth of the fields into the pipe wall
under appropriate conditions. The two distinguished ranges of $v$,
which we will refer to as ``low'' and ``high'' speed \cite{recall2},
correspond to $l_0 \gg R_1$ and $l_0 \ll R_1$, respectively
\cite{tacit}. We will first consider the low-speed regime, as it is
the relevant one in practice.  In this speed range, we have $\alpha
\simeq 1-i/({\mu}_{rel}^{-1}l_0 k)$, so that $\alpha k R_1 \simeq
kR_1 -i (R_1/{\mu}_{rel}^{-1}l_0)$, and similarly for $\alpha k
R_2$. Thus the imaginary part of the argument of various modified
Bessel functions is much smaller than the corresponding real part.
This would in turn imply that, to the leading order, the imaginary
part of those functions is linear in $R_1/l_0$ and $R_2/l_0$. This
fact implies the same for $Q$, to wit, that in the low-speed regime
the imaginary part of $Q$ is linear in $R_1/l_0$ (and $R_2/l_0$),
hence proportional to $\sigma v$ \cite{mu}. This confirms the
expectation that in the low-speed regime where $l_0 \gg R_1$, the
braking force is a linear drag proportional to the conductivity of
the pipe wall.

The linear nature of the drag force in the low-speed regime is in
evidence for all cases in Fig.~\ref{fig1}.  In typical demonstration
setups, one would expect terminal speeds of the order of $1\,\,
\textrm{m}{\textrm{s}}^{-1}$ or less, which corresponds to the
neighborhood of the origin in Fig.~\ref{fig1} where the drag force
is closely proportional to the magnet speed.   Indeed from the the
low-speed condition $\mu \sigma v R_1 \ll 1$ given above, we can
estimate the relative deviation from linearity to be of the order of
${(\mu \sigma v R_1)}^2$.  For case (d) of Fig.~\ref{fig1}, this
estimate of deviation gives ${(0.63 v[\textrm{m}
\,{\textrm{s}}^{-1}])}^2$, which amounts to about $2.5$\% at a
terminal speed of $25$ m ${\textrm{s}}^{-1}$.

In the high-speed regime (which would require the projection of the
magnet into the pipe with a suitably high initial speed) where $l_0
\ll R_1$, we have $\alpha \simeq (1-i)/{(2 {\mu}_{rel}^{-1}l_0
k)}^{\frac{1}{2}}$ for values of $k R_1 \approx 1$ which provide the
main contribution to the integral in Eq.~(\ref{25.2}). This implies
that $\alpha k R_1 \simeq (1-i){(kR_1/2)}^{\frac{1}{2}}{(R_1/
{\mu}_{rel}^{-1}l_0)}^{\frac{1}{2}}$, and similarly for $\alpha k
R_2$. Since $R_1/l_0 , R_2/l_0 \gg 1$ in this speed range, the
arguments of the modified Bessel functions in $Q$ which involve
$\alpha$ will scale as ${(R_1/l_0)}^{\frac{1}{2}}$ or
${(R_2/l_0)}^{\frac{1}{2}}$ for $k R_1 \approx 1$. Since these
functions behave exponentially for large values of their argument,
the said scaling behavior results in an overall suppression
$\text{Im}(Q)$ in the high speed regime, corresponding to a decrease
in the drag force. In other words, the drag force is a decreasing
function of $\sigma v$ in the high-speed regime. Physically, a
pronounced skin effect which suppresses the penetration of the field
into the pipe wall takes hold at high speeds, thereby reducing the
eddy currents and the drag force resulting from them. Indeed,
recalling from section \ref{introB} that the dominant time scale in
the magnet-pipe system is $R_1/v$, which corresponds to a frequency
$\omega \approx v/R_1$, we see that the high speed condition $l_0
\ll R_1 $ is equivalent to the inequality ${(\mu \sigma
\omega)}^{-\frac{1}{2}} \ll R_1$ \cite{tacit}. But this last
condition states that the skin depth corresponding to the dominant
time scale is much smaller than $R_1$ and, barring unusually
thin-walled pipes, much smaller than $R_2 - R_1$ as well. In other
words, we have ${(\mu \sigma \omega)}^{-\frac{1}{2}} \ll R_2 - R_1$
in this limit, which is the condition for a pronounced skin effect
\cite{jackson}. It is appropriate to repeat here that the high speed
limit is not likely to occur in typical arrangements of the
magnet-pipe demonstration.

The nonlinear behavior of the drag force at intermediate speeds and
its eventual drop at high speeds deduced above are clearly in
evidence for all cases displayed in Fig.~\ref{fig1}. Note that the
high-speed decline in the drag force becomes sharper as the source
currents in the magnet are more highly concentrated. Recalling our
discussion of the singular geometry above, we see the reason for
this behavior: as source currents become more concentrated,
short-distance, equivalently high-$k$, contributions become more
important, a feature that runs contrary to the high-speed situation
where high $k$ values are relatively less important.

\subsection{\label{condsus} Dependence on Conductivity and
Susceptibility}

\begin{figure}
\includegraphics[]{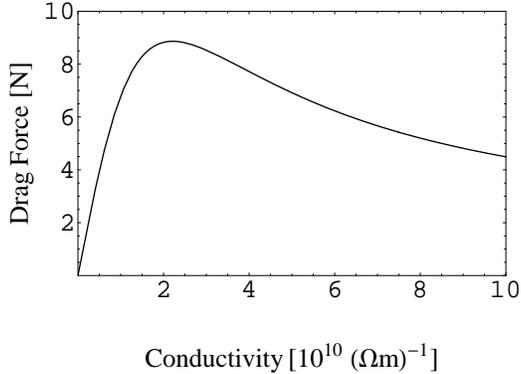}
\caption{Plot of the drag force versus the conductivity of the pipe
for case (d) with $v=0.10$ m ${\textrm{s}}^{-1}$}. \label{fig2}
\end{figure}

Recall that the drag force dependence on the magnet speed is through
the combination ${\mu} \sigma v$.  This implies that the drag force
behavior versus $\sigma$ follows the same pattern as for $v$. In
particular, higher conductivity makes for a larger drag force at low
speeds hence the use of copper tubes for demonstration purposes. At
high speeds, however, the drag force decreases with pipe
conductivity, just as it does with magnet speed. An interesting
conclusion, therefore, is that the drag force vanishes for a perfect
conductor which, as explained above, is simply a consequence of a
strong skin effect. This behavior is displayed in Fig.~\ref{fig2},
which is a plot of the drag force versus $\sigma$ for case (d) of
section \ref{shape} at a magnet speed of $0.10$ m
${\textrm{s}}^{-1}$. It is important to note that the decline of the
drag force for high values of $\sigma$ is not directly applicable to
practical arrangements which usually correspond to the low-speed (or
possibly intermediate-speed) regime . Also to be noted is the fact
that the designation ``perfect conductor'' here implies electric
conduction without resistance, and is to be distinguished from
``superconductor'' which in addition implies a distinct magnetic
behavior as noted below. Of course inasmuch as a superconductor has
zero resistance, the above argument shows that the magnetic drag
force vanishes for a superconducting pipe. Needless to say, this
conclusion as well as the one above for a perfect conductor,
directly follow from the energy conservation principle, which in
this case asserts that there can be no drag force without a
corresponding dissipation of power in the pipe.

This brings us to the question of how the magnetic properties of the
pipe wall affect the drag force \cite{murel}.  Thus far we have
assumed a linear magnetic material of relative permeability
${\mu}_{rel}$ for the pipe wall. Inasmuch as the magnetic
susceptibility of non-ferromagnetic common metals differs little
from that of the vacuum, one can set ${\mu}_{rel} \simeq 1$ for
practical purposes, as we did for the cases displayed in
Fig.~\ref{fig1}. However, it is physically interesting and
meaningful to consider the extreme cases of ${\mu}_{rel} \rightarrow
0$ and ${\mu}_{rel} \rightarrow \infty$, corresponding to perfect
diamagnetism and perfect paramagnetism, respectively.  Perfect
diamagnetic behavior is exemplified by a (type I) superconductor
which would exclude any magnetic field from its interior (save for a
very small penetration depth). This phenomenon, known as the
Meissner effect, is not a mere consequence of perfect conductivity
and serves to distinguish a superconductor from a material that
conducts electricity without dissipation \cite{ashmerm}. Perfect
paramagnetism, on the other hand, is approximately realized by
``soft'' ferromagnetic materials which have a narrow hysteresis loop
and can be idealized as highly susceptible, linear magnetic
materials.

An inspection of Eq.~(\ref{25.2}) shows that relative permeability
occurs in the quantities $\alpha$ and $\beta =\alpha / {\mu}_{rel}$
through the combination ${\mu}_{rel} {\mu}_0 \sigma v$ in $\alpha$,
and in addition, $\beta$ is inversely proportional to ${\mu}_{rel}$.
Leaving the latter aside for the moment, we conclude that the drag
force dependence on ${\mu}_{rel}$ is much like its dependence on the
magnet speed.  Thus the drag force must vanish for both ${\mu}_{rel}
\rightarrow 0$ and ${\mu}_{rel} \rightarrow \infty$, corresponding
to strongly diamagnetic and paramagnetic limits.  As further
discussed in appendix \ref{magd}, the additional dependence of
$\beta$ on ${\mu}_{rel}$ does not change these qualitative features.
Thus the drag force corresponding to a superconductor vanishes not
only because of its perfect conductivity but also due to its perfect
diamagnetism. Similarly, the drag force is seen to be small for a
strongly paramagnetic conductor, or soft ferromagnetic alloys that
behave like one.  The asymptotic behavior of the drag force for
${\mu}_{rel} \ll 1$ and ${\mu}_{rel} \gg 1$ is analyzed in
appendices \ref{magd} and \ref{magp}, respectively, where the
foregoing conclusions are explicitly confirmed.

\begin{figure}
\includegraphics[]{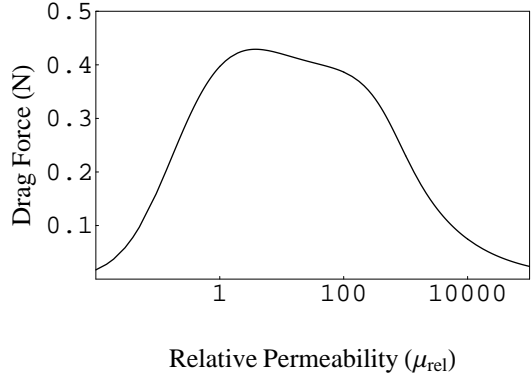}
\caption{Plot of the drag force versus the relative permeability of
the pipe for case (d) with $v=1.0$ m ${\textrm{s}}^{-1}$}.
\label{fig3}
\end{figure}

Figure \ref{fig3} shows a plot of the drag force versus
${\mu}_{rel}$ for case (d) of section \ref{shape} at a magnet speed
of $1.0$ m ${\textrm{s}}^{-1}$, where the features deduced above are
in evidence.

\section{\label{concl} Concluding Remarks}

The treatment of the magnet-pipe system in this paper has been based
on axial symmetry.  Therefore, any deviation of the magnet from a
linear, axially centered motion such as wobbling or tumbling would
violate this underlying assumption and cause a departure from the
predicted results \cite{mu}. Where necessary, the magnet can be
placed inside an electromagnetically inactive casing with an optimum
profile for stability.  This procedure would also serve to maintain
a fixed air drag coefficient when comparing magnets of different
profile.  Similarly, the analysis in this paper assumes an
infinitely long pipe, so that in practice the magnet ends must be
many times the interior pipe diameter away from the pipe ends when
measurements are taken.

The results of this paper can be used for precision magnetic braking
studies and applications.  The computer program provided for use
with this paper \cite{mathematica} has been written for the case of
uniform magnetization, but can readily be modified to deal with the
general case using Eq.~(\ref{25.1}).  For precision studies, it may
be preferable to use  an electromagnet instead of a permanent
magnet, since the former allows a more convenient characterization
of sources and fields.

\begin{acknowledgments}
We would like to thank David Jackson for reading the manuscript and
making helpful suggestions.  MHP's work was supported in part by a
research grant from California State University, Sacramento.
\end{acknowledgments}


\appendix
\section{\label{lspeed} Low Magnet Speed}
Here and in the following we will outline the development of a few
limiting expressions for the drag force given in Eq.~(\ref{25.2}).
The methods used are those of approximation and asymptotic analysis
of integrals, the details of which would take us beyond the scope of
this paper \cite{asympt}.

As a preliminary step we recall that in all cases except for
appendix \ref{magd} the main contributions to the integral of
Eq.~(\ref{25.2}) originate from the region $k \simeq {R}_{1}^{-1}$.
This fact makes it convenient to rescale the integration variable
therein according to $k=u/R_1$. Effecting this substitution, we
obtain
\begin{equation}
\mathbf{F}^{uni}=-\hat{\mathbf{v}} \frac {\mu_0 m^2 }{2
{\pi}^2{R_1}^4} {\int}_{0}^{+\infty} du u^3 {\left[\frac
{\sin(uL/2R_1)}{ (uL/2R_1)} \right]}^2 {\left[\frac
{{I}_{1}(ua/R_1)}{ (ua/2R_1)} \right]}^2
\textrm{Im}[Q(u,s/R_1,{R_1}/ l_0 ,{\mu}_{rel} )], \label{29}
\end{equation}
where we have explicitly defined the dependence of $Q$ on three
dimensionless parameters which characterize the dimensions and
material properties of the pipe as well as the magnet speed.  Note
that $s$ stands for the thickness of the pipe wall here.  We shall
use this representation to derive asymptotic expressions for the
drag force in a number of interesting limiting cases.

The low-speed regime is defined by the condition ${R_1}/l_0 \ll 1$
\cite{tacit}. To establish the fact that $\mathbf{F}^{uni}$ is
linear in $v$ in the low-speed limit, as discussed in Sec.
\ref{speed}, we note that $Q(u,s/R_1,0,{\mu}_{rel} )$ is real at
$v=0$ (corresponding to the vanishing of the drag force at zero
speed), so that an expansion of $\textrm{Im}(Q)$ about $v=0$ using
the asymptotic properties of the modified Bessel functions
\cite{arfken} yields $\textrm{Im}[Q(u,s/R_1,{R_1}/l_0 ,{\mu}_{rel}
)] \cong \mathcal{L} {R_1}/l_0$ for small ${R_1}/l_0 $.  The
quantity $\mathcal{L}$ equals $\textrm{Im}[\partial
Q(u,s/R_1,0,{\mu}_{rel} )/ \partial ({R_1}/l_0)]$, and is given by a
long expression which need not be reproduced here. Substituting the
approximate form of $\textrm{Im(Q)}$ in Eq.~(\ref{29}), we
immediately obtain the linear drag behavior in the low-speed limit:
\begin{equation}
{\mathbf{F}}^{lsp}\cong - \mathcal{C} \sigma v \hat{\mathbf{v}}
\,\,\, ({\mu}_{rel} {\mu}_0 \sigma v R_1 \ll 1), \label{30}
\end{equation}
where $\mathcal{C}$ depends on all parameters other than $\sigma$
and $v$.

\section{\label{idl} The Idealized Model}

The idealized model involves three assumptions: (a) the point-dipole
limit, $L/R_1 \rightarrow 0, \, a/R_1 \rightarrow 0$, (b) the
low-speed approximation, ${R_1}/l_0 \ll 1$, and (c) the thin-wall
approximation, $s/R_1 \ll 1$. The point-dipole limit is readily
implemented by setting the two shape factors equal to unity.  Items
(b) and (c), on the other hand, require applying the thin-wall
approximation to the quantity $\mathcal{C}$ introduced in
Eq.~(\ref{30}). Using the properties of the modified Bessel
functions, we find from Eqs.~(\ref{19}-\ref{21}) that ${\mathcal{C}}
\rightarrow (s/R_1) u [{K}_{1} (u)]^2 $. Therefore,
\begin{equation}
{Q}^{idl}=i{\mu}_0 (s/R_1) u [{K}_{1} (u)]^2 \sigma v, \label{31}
\end{equation}
where ${Q}^{idl}$ represents $Q$ in the ideal limit. Substituting
this result in Eq.~(\ref{29}) (with the shape factors set to unity),
we find
\begin{equation}
{\mathbf{F}}^{idl}=- \frac{ 45 {\mu}_{0}^{2} {m}^{2}s }{1024
{R_1}^{4}}{\sigma}{v} \hat{\mathbf{v}}, \label{32}
\end{equation}
in agreement with previous results \cite{saslow}.

\section{\label{high} High Magnet Speed}

As stated in Sec. \ref{speed}, in the high-speed regime where $l_0
/R_1 \ll 1$, the quantity $\alpha$ in Eqs.~(\ref{19}-\ref{21}) has a
large magnitude which forces the respective modified Bessel
functions to their asymptotic range.  Since the four quantities
$T_{ij}$ in Eqs.~(\ref{20}) have the same asymptotic limit
\cite{arfken}, they cancel out of Eq.~(\ref{19}). The remaining
terms can then be simplified using the properties of the modified
Bessel functions, leaving the result
\begin{equation}
\textrm{Im}({Q})\cong \frac{1}{k R_1}
\frac{-\textrm{Im}(\beta)}{{[{I}_{0} (|k| R_1)]}^{2}+|\beta|^2
{[{I}_{1} (|k| R_1)]}^{2}} \,\,\, ({\mu}_{rel} {\mu}_{0} \sigma v
\gg R_1). \label{33}
\end{equation}
The remaining $v$ dependence in Eq.~\ref{33} resides in $\beta
=\alpha / {\mu}_{rel}$.  Recalling from Sec. \ref{speed} that
$\alpha \rightarrow (1-i)/{(2 {\mu}_{rel}^{-1}l_0 k)}^{\frac{1}{2}}$
in the high-speed limit, we can reduce the above equation to
\cite{tacit}
\begin{eqnarray}
\textrm{Im}({Q})&\cong& \frac{ \sqrt{{\mu}_{rel}}} {\sqrt{2 k
{\mu}_{0} \sigma v }R_1 {[{I}_{1}(|k|R_1)]}^{2}}  \nonumber \\
&=& \frac{\sqrt{ {\mu}_{rel} }} {\sqrt{2 u R_1/ l_0}
{[{I}_{1}(u)]}^{2}} \,\,\, ({\mu}_{rel} {\mu}_{0} \sigma v \gg R_1).
\label{34}
\end{eqnarray}
Upon substituting this expression in Eq.~\ref{29}, we find
\begin{equation}
{\mathbf{F}}^{hsp}=-  \frac {0.274  m^2 }{{R}_{1}^{9/2}}
{\mathcal{F}}_1 (a/R_1,L/R_1)\sqrt{ \frac{{\mu}}{{\sigma v}
}}\hat{\mathbf{v}}, \label{35}
\end{equation}
where ${\mathcal{F}}_1$ is a form factor which depends on the
dimensions of the magnet as fractions of the interior diameter of
the pipe. It is defined by
\begin{equation}
{\mathcal{F}}_1 (a/{R_1},L/{R_1}) = \frac{{{\mathcal{G}}_1
}(a/{R_1},L/{R_1})} {{\mathcal{G}}_1 (0,0)}, \label{36}
\end{equation}
where
\begin{equation}
{\mathcal{G}}_1 (a/R_1,L/R_1)={\int}_{0}^{+\infty} du
{u}^{5/2}{\left[\frac {\sin(uL/2R_1)}{ (uL/2R_1)} \right]}^2
{\left[\frac {{I}_{1}(ua/R_1)}{ (ua/2R_1)} \right]}^2
{[{I}_{1}(u)]}^{-2}. \label{37}
\end{equation}
Note that the form factor ${\mathcal{F}}_1$ has been normalized to
unity at the point-dipole limit.

The asymptotic formula in Eq.~(\ref{35}) is valid in the high-speed
limit ${\mu}_{rel} {\mu}_{0} \sigma v \gg R_1$ and describes the
behavior of the drag force for very high magnet speed or pipe wall
conductivity \cite{tacit}.

\section{\label{magd} Highly Diamagnetic Pipe}

This is the limit $\mu_{rel} \rightarrow 0$, which is appropriate to
a highly diamagnetic pipe wall material.  In this limit the factor
$\beta$, which occurs in $Q$, grows large and severely suppresses
the magnitude of $\textrm{Im}(Q)$ in Eq.~(\ref{29}), and in addition
limits its contributions to very small values of $u$. This
mathematical behavior reflects the physical phenomenon of magnetic
flux expulsion that accompanies the approach to perfect
diamagnetism.

Since only small values of $u$ are important in Eq.~(\ref{29}), we
can replace all modified Bessel functions by their asymptotic values
in the small argument limit \cite{arfken}. This leads to the
following approximate expression for $Q$:
\begin{equation}
Q =-\frac{{\alpha}^{2} \ln(R_2/R_1) }{{\mu}_{rel} +\frac{1}{2} u^2
\ln(R_2/R_1){\alpha}^{2} }. \,\,\,  (u \ll 1) \label{38}
\end{equation}
Next, we find $\textrm{Im}(Q)$ and change the variable of
integration in Eq.~(\ref{29}) according to
$w={\mu}_{rel}^{-\frac{1}{2}} {\ln(R_2/R_1)}^{\frac{1}{2}}\, u$.
After some calculation and simplification, we find
\begin{equation}
{\mathbf{F}}^{hdm}=-\mathcal{K} {\int}_{0}^{+\infty} d \zeta
\,{\zeta}^2 {[1+\frac{1}{2} {\zeta}^2]}^{-2} \hat{\mathbf{v}},
\label{39}
\end{equation}
where
\begin{equation}
\mathcal{K}= \frac {{\mu}_{0}^{2} m^2  \sigma v }{2 {\pi}^2
{R}_{1}^{3}} {[\ln(R_2/R_1)]}^{-\frac{1}{2}} {\mu}_{rel}^{3/2}.
\label{40}
\end{equation}
Finally, we carry out the integral in Eq.~(\ref{39}) to arrive at
the asymptotic behavior of the drag force:
\begin{equation}
{\mathbf{F}}^{hdm}=-\frac {{\mu}_{0}^{2} m^2  \sigma v }{2
{\pi}\sqrt{2} {R}_{1}^{3}} {[\ln(R_2/R_1)]}^{-\frac{1}{2}}
{\mu}_{rel}^{3/2} \, \hat{\mathbf{v}}. \label{41}
\end{equation}

The above formula is valid in the limit ${\mu}_{rel} \rightarrow 0$,
and is appropriate to a highly diamagnetic pipe.

\section{\label{magp} Highly Paramagnetic Pipe }

Here we are considering the limit $\mu_{rel} \rightarrow \infty$
appropriate to a highly susceptible pipe wall material\cite{murel}.
Recall from \ref{condsus} that the dependence of $Q$ on
${\mu}_{rel}$ occurs through the combination ${\mu}_{rel} {\mu}_0
\sigma v$ in $\alpha$, and in addition through $\beta =\alpha /
{\mu}_{rel}$. Consequently, Eq.~\ref{33} which is appropriate in the
high-speed regime holds here as well.  Recalling further that $\beta
\rightarrow (1-i)/{(2 {\mu}_{rel}l_0 k)}^{\frac{1}{2}}$ in this
limit, we find
\begin{equation}
\textrm{Im}({Q})\cong \frac{1} {\sqrt{2 u^3 l_0 /R_1 }
{[{I}_{0}(u)]}^{2} \sqrt{ {\mu}_{rel} }} \,\,\, ({\mu}_{rel}
 \gg 1). \label{42}
\end{equation}
At this point we follow the steps subsequent to Eq.~(\ref{33}) to
find
\begin{equation}
{\mathbf{F}}^{hpm}=-  \frac {0.0536 {\mu}_{0}^{2} m^2
}{{R}_{1}^{7/2}} {\mathcal{F}}_0 (a/R_1,L/R_1)\sqrt{ \frac{{\sigma
v} }{{\mu}}}\hat{\mathbf{v}}, \label{43}
\end{equation}
where ${\mathcal{F}}_0$ is a form factor which depends on the
dimensions of the magnet as fractions of the interior diameter of
the pipe. It is defined by
\begin{equation}
{\mathcal{F}}_0 (a/{R_1},L/{R_1}) = \frac{{{\mathcal{G}}_0
}(a/{R_1},L/{R_1})} {{\mathcal{G}}_0 (0,0)}, \label{44}
\end{equation}
where
\begin{equation}
{\mathcal{G}}_0 (a/R_1,L/R_1)={\int}_{0}^{+\infty} du
{u}^{3/2}{\left[\frac {\sin(uL/2R_1)}{ (uL/2R_1)} \right]}^2
{\left[\frac {{I}_{1}(ua/R_1)}{ (ua/2R_1)} \right]}^2
{[{I}_{0}(u)]}^{-2}. \label{45}
\end{equation}
Note that the form factor ${\mathcal{F}}_0$ has been normalized to
unity at the point-dipole limit.

The asymptotic formula given in Eq.~(\ref{42}) is valid in the limit
${\mu}_{rel}  \gg 1$ and describes the behavior of the drag force
for a highly susceptible pipe.


\begin{thebibliography}{}
\bibitem{si} The SI system of units will be used throughout this paper.
\bibitem{magdip} The dipole radiation can be found from the magnetic dipole
radiation formula, $({\mu}_{0}/6 \pi
{c}^{3}){\ddot{\mathsf{M}}}^{2}$, where $\mathsf{M}$ is the dipole
moment in question. Using this formula, and recalling that the
magnet's dipole moment in the laboratory frame equals $m/\gamma$,
we find the result $({{\mu}_{0}{m}^{2} / 6 \pi {c}^{7}})
{(\dot{v}^{2}+ v \ddot{v})}^2$. This quantity is in general
smaller than the quadrupole radiation by a factor of order
${(v/c)}^2$.
\bibitem{region0} There is a zeroeth region, ($0$), defined by $0 \leq \rho \leq
a$, which we have not included here.  The potential in this region
is obtained from that in region (i) by substituting the appropriate
representation of ${\tilde{A}}^{(i)}(\rho,k)$ valid for $0 \leq \rho
\leq a$. The latter is defined in the paragraph following
Eq.~(\ref{6}).
\bibitem{Qhas} The imaginary part of Q is positive, a
fact that is not readily obvious from its definition.
\bibitem{scale} An analogus situation obtains for the
electrostatic force exerted on a charged object by its image with
respect to a conducting or (linear) dielectric medium.  Such a
force would scale with the square of the real charge.
\bibitem{thinwall} In effect the thin-wall approximation assumes
uniform fields and eddy currents within the pipe wall thus ignoring
the inevitable fall-off of all disturbances away from the axis. This
is the reason why the thin-wall approximation in general
overestimates the drag force.
\bibitem{vac} Air drag effects can of course be eliminated by
placing the pipe-magnet system in an evacuated chamber.
\bibitem{recall2} Recall that we have already restricted $v$ to
nonrelativistic values.
\bibitem{tacit} Here we are implicitly assuming that ${\mu}_{rel}$
is of the order of unity.  We will consider departures from this
assumption in \ref{condsus}, and more specifically, the possibilites
of ${\mu}_{rel} \ll 1$ and ${\mu}_{rel} \gg 1$ in appendices
\ref{magd} and \ref{magp}, respectively.
\bibitem{mu} Because the dependence of $Q$ on $\mu$ is not just through
the Bessel functions, this argument fails to give the full
dependence of the force on $\mu$.
\bibitem{green} This is essentially the method of Green's function for
solving boundary value problems associated with linear systems; see,
e.g., the textbook by Arfken and Weber \cite{arfken}, or Jackson
\cite{jackson}, for a detailed description of this method.
\end{thebibliography}
{}

\end{document}